\documentstyle[11pt,aaspp4]{article}

\lefthead{Cocchi et al.}
\righthead{X-ray bursts from {\em 1RXS J170930.2-263927}}

\begin{document}

\title{X-ray bursts from 1RXS J170930.2-263927 = XTE J1709-267}

\author{M. Cocchi\altaffilmark{1}, A. Bazzano, L. Natalucci and P. Ubertini}
          \affil{Istituto di Astrofisica Spaziale {\em(IAS/CNR)}, 
           via Fosso del Cavaliere, 00133 Roma, Italy} 
\author{J. Heise, J.M. Muller\altaffilmark{2}, M.J.S. Smith\altaffilmark{2} 
            and J.J.M in 't Zand}
          \affil{Space Research Organization Netherlands {\em (SRON)}, 
           Sorbonnelaan 2, 3584 TA Utrecht, The Netherlands}


\altaffiltext{1}{e-mail address: wood@ias.rm.cnr.it}
\altaffiltext{2}{also {\em BeppoSAX} Science Data Centre, Nuova Telespazio, 
                      via Corcolle 19, 00131 Roma, Italy}


\begin{abstract}
We report evidence of type-I X-ray bursting activity from the transient
{\em 1RXS J170930.2-263927} 
when it was in outburst early 1997.  This identifies the source as a probable 
low-mass X-ray binary containing a neutron star.  The error boxes of the 
detected bursts and of the persistent emission, as obtained with the 
{\em Wide Field Cameras} on board the {\em BeppoSAX} satellite, rule out an 
association with the proposed radio counterpart (\cite{Hjel97}).
\end{abstract}


\keywords{binaries: close, individual ({\em 1RXS J170930.2-263927})
          --- X-rays: bursts}


%

\section{Introduction}

{\em 1RXS J170930.2-263927} was first observed in hard X-rays with
the {\em Proportional Counter Array (PCA)} on board the {\em Rossi 
X-ray Timing Explorer (RXTE)} (\cite{Mars97}). The source, reported as a new
{\em XTE} transient source ({\em XTE J1709-267}), was detected for about 20~s during 
two different satellite manoeuvres, on 1997 January~16.016 and 1997 January~18.147 
respectively.
The uncertainty of the source position was estimated to be 6', 
suggesting a possible association of {\em XTE J1709-267} with the soft X-ray source 
{\em 1RXS J170930.2-263927} included in the {\em ROSAT All-Sky Survey} catalogue (\cite{Vogs96}), 
5' from the {\em PCA} centroid position (see Figure 3). \\
A possible radio counterpart of the source was found in the 1.42 and 4.9 GHz bands 
(\cite{Hjel97}).
The radio position, given with better than 1" uncertainty, is $2\farcm 8$ from the 
{\em PCA} centroid position and 6' from the {\em ROSAT} position.  
So the hard X-ray source could only be associated either with 
{\em 1RXS J170930.2-263927} or with the proposed radio counterpart, but not with both. \\
The X-ray characteristics of the {\em RXTE} source are not known in detail. 
The spectrum obtained by the {\em PCA} could be fitted with a power law 
function\footnote{
i.e. n(E) $\propto$ E$^{-\Gamma}$, where n(E) is the number of 
photons with energy E and $\Gamma$ is the photon index.
}
with a $3.06\pm0.06$ photon index or by a thermal 
bremsstrahlung model with {\em k}T= $4.7\pm0.2$ keV. 
In both cases the column density was found in excess of 
$2\times10^{22}$~cm$^{-2}$, thus indicating the source to be close to the Galactic 
Center.  
During the PCA observations the source had an intensity of 140 mCrab in the
2-10 keV band, while the 1-2.4 keV flux of {\em 1RXS J170930.2-263927} at its 
detection time (spring 1996) was only 13 mCrab. \\
After its detection in hard X-rays, {\em 1RXS J170930.2-263927} was monitored by the 
{\em All Sky Monitor} ({\em ASM}) on board {\em RXTE} in the 2-10 keV range. 
The source was still bright (140-200 mCrab) up to February~20, then declined to 
80-120 mCrab (March~12-18) and to 25-65 mCrab since March~19. 
The source was no longer in outburst since April~7, its daily averaged flux being 
below 13 mCrab. \\
The {\em RXTE-ASM} lightcurve of the outburst is displayed in Figure 1.
The lightcurve does not show uncommon features when compared with the average lightcurves of 
the X-ray novae.  Following the morphological classification proposed by \cite{Chen97}, we can 
associate the source with the fast-rise exponential-decay class, even if the decay looks 
almost linear and perhaps dropping faster about 50 days after the maximum.  
The {\em ASM} peak luminosity of the outburst is 0.21 Crab, a rather low but not uncommon value 
among the X-ray novae, where selection effects in the studied samples are not 
negligible.  
We can derive that at the outburst epoch the source intensity increased  
by more than an order of magnitude.

In the next Section we summarize the {\em BeppoSAX-WFC} telescope characteristics and
report on the observations of {\em 1RXS J170930.2-263927}.  In Section~3 the 
performed spectral and time analysis are presented, while in Section~4 the 
implications of our results on the knowledge of the source characteristics are 
briefly discussed.  In particular, the association of {\em 1RXS J170930.2-263927} 
with the class of transient neutron-star low-mass X-ray binaries is proposed.

\begin{figure}
\plotone{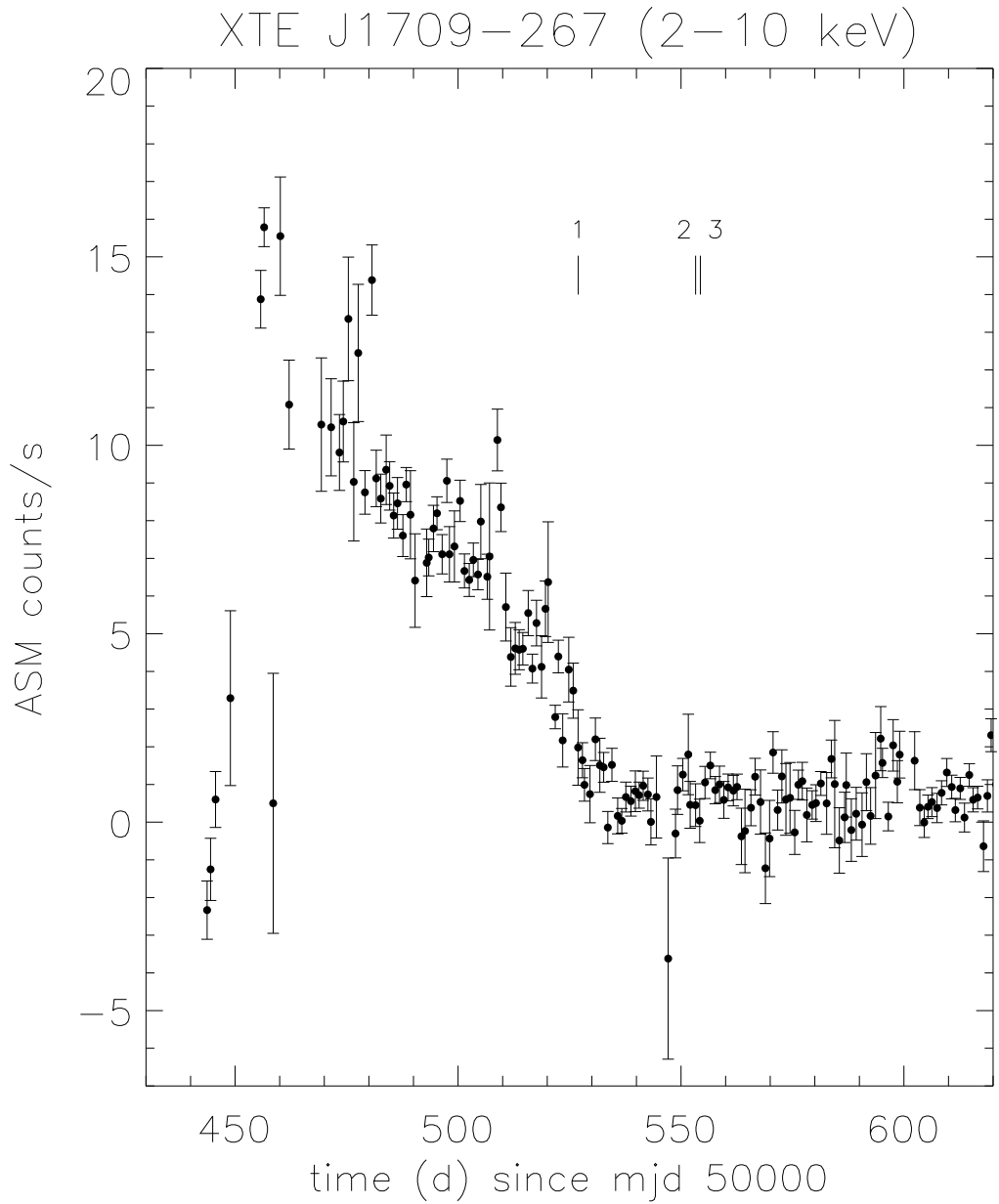}
\caption{
     {\em RXTE-ASM} lightcurve of {\em 1RXS~J170930.2-263927}.
     The markers 1,2,3 indicate the epochs of the observed {\em WFC} bursts.
   \label{fig1}}
\end{figure}

\placefigure{fig1}

\section{Observations}

The {\em Wide Field Cameras (WFC)} on board the {\em BeppoSAX} satellite consist
of two identical coded mask telescopes (\cite{Jage97}).
The two cameras point in opposite directions each covering a $40\arcdeg\times40\arcdeg$
field of view.  This field of view is the largest ever flown for an X-ray 
imaging device.
With its source location accuracy better than $0\farcm 6$ (68\% confidence level),
a time resolution of 0.244 ms, and an energy resolution of 18\% at 6 keV, the
{\em WFC}s are very effective in studying hard X-ray (2-28 keV) transient
phenomena. The imaging capability, combined with the good instrument sensitivity
(a few mCrab in $10^{4}$ s), allows an accurate monitoring of complex sky 
regions, like the Galactic Bulge.\\
The data of the two cameras are systematically searched for bursts and flares by 
analyzing the time profiles of the detectors in the 2-11 
keV energy range with a 1~s time resolution.  Reconstructed sky images are
generated for any statistically meaningful enhancement, to
identify possible bursters. The accuracy of the reconstructed position, which 
of course depends on the burst intensity, is typically better than 5'.
This analysis procedure demonstrated its effectiveness throughout the Galactic Bulge
{\em WFC} monitoring (\cite{Cocc97a}), leading to the identification of $\sim 530$
 X-ray bursts, 156 of which from the {\em Bursting Pulsar GRO J1744-28}, in a 
total of about 1100 ks observing time (the exposure time values are corrected for all
dead times due to Earth occultation and South Atlantic Anomaly).  
A total of 12 new X-ray bursting sources were found.

{\em 1RXS J170930.2-263927} is in the the field of view whenever the {\em WFC}s point at the 
Galactic Centre region, being $9\fdg 4$ away from the Sgr A position. 
The Galactic Bulge field was observed for a total of about
1100 ks, the monitoring being divided in 4 observation campaigns during 1996 through 1998.
Unfortunately, {\em 1RXS J170930.2-263927} was out of visibility at the
epoch of the outburst onset (January '97), so only data from the declining phase and 
quiescence are available.

During a 102.6 ks observation started on 1997 March~18.04 
a burst was detected at a position consistent with that of {\em 1RXS J170930.2-263927} 
at March~18.21058  {\em UT} (\cite{Cocc97b}). The event occurred during the egress from 
the Earth's occultation. 
Two more bursts were detected from the same position in a different observation 
shift that started on April~13.16 and lasting for 97.8 ks. The time 
profiles of the three bursts are plotted in Figure 2.
The displayed profiles of the bursts are detector profiles constructed by integrating 
the photons in the source-illuminated part of the detector. This improves the signal to 
noise ratio of the event.  The background is the sum of the diffuse X-ray background, the
particles background and the contamination of other sources in the field of view.
Source contamination is the dominating background component for crowded fields like the 
Galactic Bulge.
Neverthless, the probability of source confusion during a short time scale event like an
X-ray burst is negligible.
 Details on the three observed bursts are summarized in Table 1. \\
Burst searches from {\em 1RXS J170930.2-263927} were performed on all the data available from the
'96-'98 {\em BeppoSAX-WFC} Galactic Bulge monitoring campaigns but no other events were detected.

\begin{figure}
\plotone{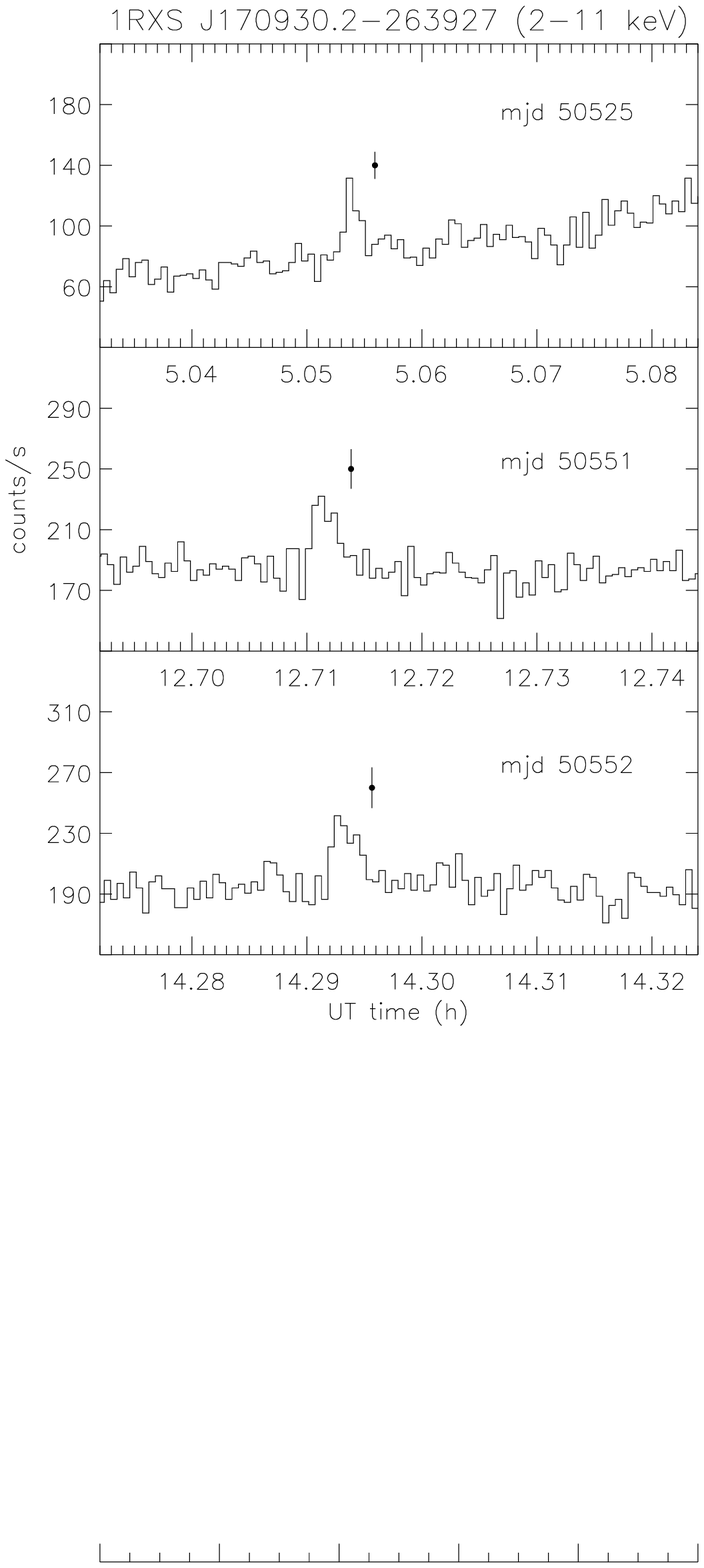}
\caption{
      2-11 keV time profiles of the observed bursts. 
    \label{fig2}}
\end{figure}

\placefigure{fig2}
\placetable{tbl1}




\section{Data Analysis}

The steady emission of {\em 1RXS J170930.2-263927} during its declining 
phase in March '97 has been investigated.
We analyzed data from two observations which started on March~13.52 and March~18.04 (U.T.)
for a total of 170.6 ks exposure time. \\
The intensity of the source was found to be fairly constant throughout the two 
observations, the average flux being about 28 mCrab.  The data was split in five data 
subsets in order to investigate possible spectral variability on 1 day timescale.  
The fitted parameters over all the subsets are consistent with a constant spectrum,
thus excluding spectral variability during the involved epochs.
Two spectral models were applied to the average spectrum over all the five subsets:
an absorbed power law function and a thermal bremsstrahlung spectrum.
The calculated parameters, which are consistent with the ones reported by 
{\em RXTE-PCA}, are shown in Table 2.

An absorbed blackbody model was adopted to fit the spectra of the three bursts.
To better constrain the fit, the $N_{\rm H}$ parameter was kept 
fixed ($N_{\rm H}$ = $2\times10^{22}$~cm$^{-2}$) according to what was obtained from 
the spectral analysis of the steady emission by using {\em WFC} and {\em RXTE-PCA} data.  
Moreover, two time-resolved spectra were obtained for the 2nd burst, i.e. a "peak spectrum" 
(data from the first 6~s of the burst) and a "tail spectrum" (the next 9~s of burst data).
A summary of the spectral parameters of the three bursts is given in Table 1. \\
Blackbody spectra allow to determine the relationship between the average radius of the emitting
sphere $R_{\rm km}$ (in units of km) and the source distance $d_{\rm 10~kpc}$ (in units of 10 kpc).
In Table 1 the values of the $R_{\rm km}$/$d_{\rm 10~kpc}$ ratios are given, assuming isotropic
emission and not correcting for gravitational redshift and conversion to true blackbody
temperature from color temperature (see \cite{Lewi93} for details).

The calculated {\em WFC} centroid position of the observed hard X-ray source was found to 
be consistent with that of the soft {\em ROSAT} source {\em 1RXS J170930.2-263927} 
(\cite{Cocc97a}).  
In Figure 3 a sky map is shown with the error circles of the position we calculated for 
the March~18 persistent emission and for the three bursts. 
The {\em WFC} position is almost coincident with the {\em ROSAT} one, thus associating the 
hard X-ray {\em XTE} source with the soft X-ray {\em ROSAT} source {\em 1RXS J170930.2-263927}.
A possible association of both the soft and the hard X-ray sources with the proposed radio 
counterpart, which lies more than 5' away, is ruled out.

\begin{figure}
\plotone{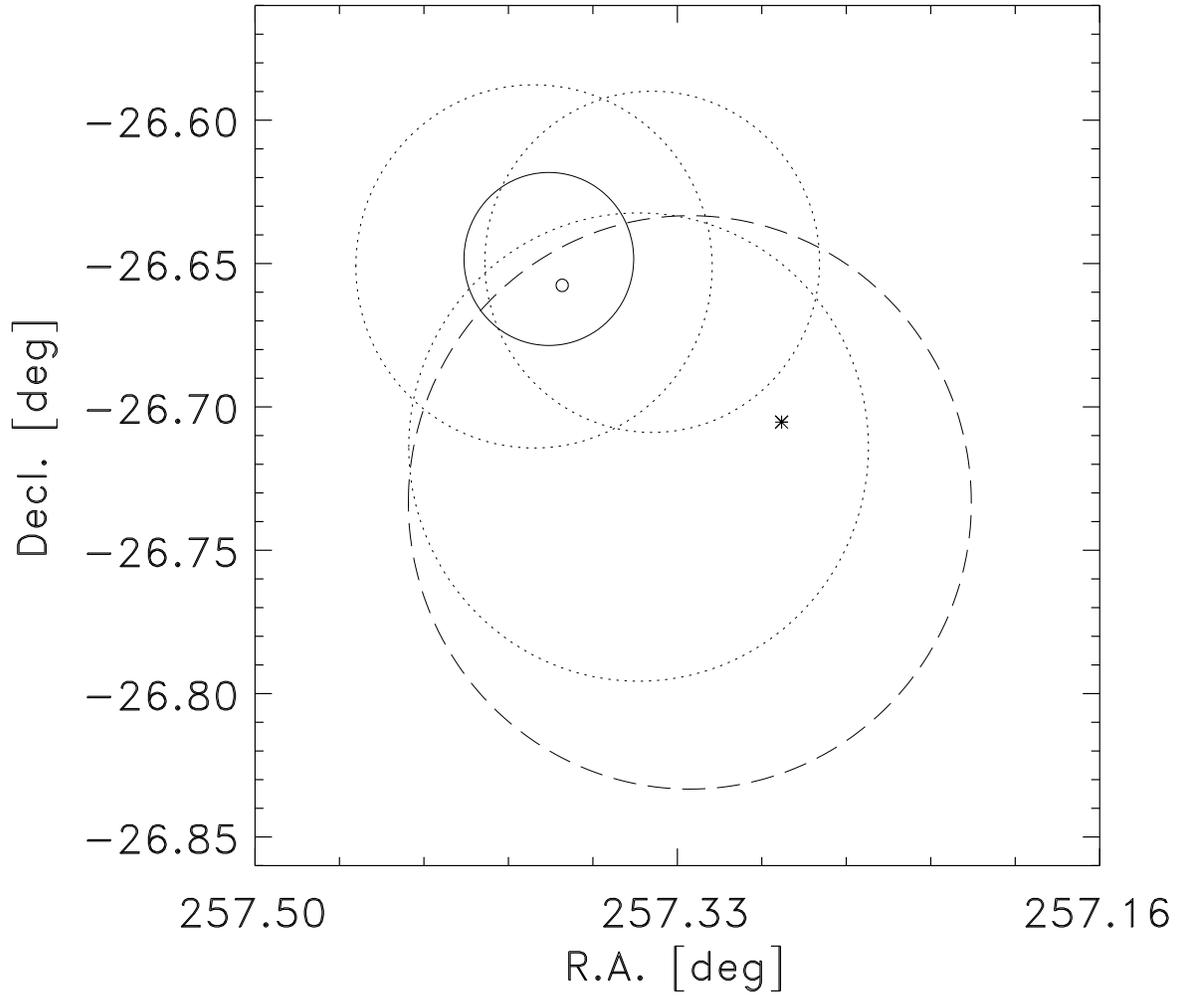}
\caption{
     A skymap of relevant error regions.
     Small solid circle: {\em 1RXS~J170930.2-263927} error circle (Voges et al. 1996).
     Large solid circle: {\em WFC-SAX} steady source error circle (99\% confidence).
     Dotted circles: error circles of the three {\em WFC}-detected bursts (99\% confidence).
     Large dashed circle: {\em XTE~J1709-267} error circle (only systematic errors, Marshall et al. 1997).
     Asterisk: position of the suspected radio counterpart (Hjellming \& Rupen 1997).
   \label{fig3}}
\end{figure}

\placefigure{fig3}
\placetable{tbl2}

\section{Discussion}

Up to date, transient phenomena like hard X-ray bursts with 
typical durations less than a few minutes are classified in two main types 
(\cite{Hoff78}). \\
Type-I bursts originate from thermonuclear flashes onto a neutron star surface 
(\cite{Woos76,Joss78,Taam79}).
All the type-I bursting sources are associated with low mass binary systems (LMXBs), following
the identification of their optical counterparts or, indirectly, from spectral or 
stellar population (e.g. globular cluster sources) characteristics (\cite{vanP95}).
The observed spectra of the bursts are typically 
well fitted by thermal blackbody emission models having temperatures {\em k}T~$\lesssim$3~keV 
(\cite{Swan77}).
The time profiles of type-I bursts 
show fast rise times, ranging from less than 1~s up to 10~s, and longer decay
times, ranging from seconds to minutes. They strongly depend on photon energy,
decays being shorter at higher energies, so that bursts appear to soften 
during the decay. This softening has been suggested to originate from the 
cooling of the neutron star photosphere (e.g. \cite{Lewi95}). \\
Type-II bursts have been observed only in two peculiar cases, namely the
{\em Rapid Burster} {\em MXB 1730-335} (\cite{Lewi76}) and the {\em Bursting Pulsar} 
{\em GRO J1744-28} (\cite{Kouv96}).
They show no spectral softening and are frequently repeated in time, intervals ranging from a few 
seconds to hours.  Blackbody fits to the type-II bursts are unacceptably poor.  
Comptonized models for {\em MXB 1730-335} (\cite{Stel88}) and a power law function 
model for {\em GRO J1744-28} (\cite{Gile96}) fit the data better.  Moreover,
the {\em Bursting Pulsar} spectra of the steady emission and the bursts are very similar.
Type-II bursts are thought to originate from accretion instabilities
(Lewin et al. 1995).

On the basis of their spectral properties and time profiles, we interpret the three 
bursts detected from {\em 1RXS J170930.2-263927} as type-I bursts. 
Even though the statistical quality of the data is limited, the blackbody emission
and the measured color temperatures are consistent with type-I bursts.
Spectral softening is marginally observed in time resolved spectra of burst 2 and 
energy resolved time profiles of all the bursts (see Table 1). 
Average decay times of $8\pm2$~s (2-6 keV) and $4\pm1$~s (6-18 keV) are obtained. \\
Type-I bursts strongly suggest a neutron star nature for the binary system. This indicates
{\em 1RXS J170930.2-263927} as a transient neutron-star LMXB. \\
In principle, it is possible that {\em 1RXS J170930.2-263927} had bursting behaviour
even before the January 1997 outburst, as the bursts occurrence doesn't seem to be 
related to the persistent flux (bursts 2 and 3 occurred when the steady flux had dropped 
by a factor $\gtrsim 10$ with respect to the time of burst 1, see Table 1).
On the other hand, it is possible that the outburst itself triggered the type-I burst mechanism, 
due to an increased accretion rate from the companion star and that the physical conditions 
that activated the bursts persisted even after the source entered a quiescent phase.  
This scenario seems likely since no bursting activity was detected from {\em 1RXS J170930.2-263927}
in the autumn '96 and autumn '97 observation campaigns.

Bursting activity from LMXB transients have already been reported in about 10 cases 
(e.g. {\em Rapid Burster, Aql X-1, Cen X-4, 0748-673, 1658-298, SAXJ 1808.4-3658}, see
Hoffman et al. 1978, \cite{Tana96}, Lewin et al. 1995, and references therein, \cite{Zand98}), thus
indicating the sources to be neutron-star binaries.
Most ($\simeq 85\%$) of the LMXB harbouring a neutron star are persistent sources, while  
all the black hole candidates in low-mass systems are transient sources (\cite{vanP95}).
Among the LMXB transients a quite large fraction of sources ($\simeq 30\%$, 
according to Chen et al. 1997, $\simeq 45\%$, according to \cite{Tana96}) are neutron-star systems,
the rest being black hole binaries.
  The population of black hole LMXB could be overstimated, since most
of them are suggested as black hole candidates on the basis of their spectral characteristics only.
Actually, for only 7 out of about 40 known transient LMXB the available mass functions suggest 
black hole systems (Chen et al. 1997). \\
Thus, the detection of type-I X-ray burst in X-ray novae can help to better constrain the populations.

Due to the uncertainty in the distance of the source, interesting parameters such as the 
luminosity of {\em 1RXS J170930.2-263927} both in outburst and in quiescence, the blackbody 
emitting region radius during the bursts and the peak luminosities of the bursts cannot be 
satisfactorily constrained. 
The interpolated $N_{\rm H}$ value computed at the {\em ROSAT} given position is $2.3\times10^{21}$~cm$^{-2}$
(\cite{Dick90}), about an order of magnitude lower than the measured value, thus suggesting 
intrinsic absorption.  This makes it difficult to use the observed $N_{\rm H}$ as a distance 
indicator. \\
A rough upper limit for the source distance can be obtained assuming the observed bursts had luminosities
below the Eddington limit ($L_{\rm Edd} \simeq 2\times10^{38}$~erg~s$^{-1}$).  Low blackbody temperatures 
($<2.5$ keV) and no evidence of double-peaked profiles (a clue of photospheric radius expansion) support 
this assumption for type-I bursts (Lewin et al. 1995).
Correcting for interstellar absorption, for burst 2 we obtain a bolometric flux of 
$L_{\rm b} = 1.6\times10^{-8}$~erg~s$^{-1}$cm$^{-2}$.  Assuming $L_{\rm b} \leq L_{\rm Edd}$, this leads to a 
maximum distance of $10\pm 1$ kpc, but a lower distance value is likely since the source is 
at relatively high ($7\fdg 9$) galactic latitude. 
We then derive upper limits of $\sim 8\times10^{37}$~erg~s$^{-1}$
for the outburst peak luminosity and $\sim 1\times10^{36}$~erg~s$^{-1}$ for the persistent
emission at the bursts 2,3 epochs.  An upper limit of $\sim 10$ km is also obtained for the average 
radius of the blackbody emitting region during the observed bursts, a value supporting the neutron
star identification.

%
%

\acknowledgments

We thank the staff of the {\em BeppoSAX Science Operation Centre} and {\em Science
Data Centre} for their help in carrying out and processing the {\em WFC} Galactic Centre
observations. The {\em BeppoSAX} satellite is a joint Italian and Dutch program.
M.C., A.B., L.N. and P.U. thank Agenzia Spaziale Nazionale ({\em ASI}) for grant support.

\clearpage

\clearpage
 
\begin{deluxetable}{crrrrrrrrrrr}
\footnotesize
\tablecaption{Summary of the {\em XTE J1709-267} bursts characteristics} \label{tbl1}
\tablewidth{0pt}
\tablehead{
\colhead{} & \colhead{burst 1}  & \colhead{burst 2}  & \colhead{burst 3}
} 
\startdata
Burst time  & March $18.21058$  &  April $13.52963$  &  April $14.59554$  \\
Burst {\em e}-folding time (2-11 keV)  & $5.2 \pm 1.8$ s & $ 7.2 \pm 2.8$ s & $ 10.0 \pm 3.5$ s \\
Maximum burst intensity flux (2-11 keV) & $0.48 \pm 0.13$ Crab & $0.50 \pm 0.12$ Crab & 
                                          $0.48 \pm 0.13$ Crab \\
Steady intensity flux (2-10 keV)    & $26 \pm 1$ mCrab & $<3$ mCrab\tablenotemark{a} & 
                                      $<3$ mCrab\tablenotemark{a} \\
\\
Burst {\em e}-folding time (2-6 keV)  & $6.6 \pm 3.2$ & $ 7.5 \pm 3.5$ s & $ 11.9 \pm 4.8$ s \\
Burst {\em e}-folding time (6-18 keV)  & $2.6 \pm 1.4$ & $ 5.5 \pm 2.8$ s & $ 7.1 \pm 4.2$ s \\
\\
Blackbody temperature (keV)   &  $2.10 \pm 0.16$  &  $1.79 \pm 0.16$  &  $1.73 \pm 0.16$  \\
Reduced $\chi^{2}$~\tablenotemark{b}     &  $0.93$  &  $0.65$  &  $0.91$  \\
{\em peak} blackbody {\em k}T (keV)     &  $-$  &  $1.83 \pm 0.26$  & $-$  \\
{\em tail} blackbody {\em k}T (keV)    &  $-$  &  $1.69 \pm 0.28$  & $-$  \\
\\
$R_{\rm km}/d_{\rm 10~kpc}$   & $7.9 \pm 2.3$  &  $11.0 \pm 3.3$  &  $11.6 \pm 3.7$  \\
\enddata

 
\tablenotetext{a}{$3\sigma$ upper limit}
\tablenotetext{b}{26 d.o.f.}

\end{deluxetable}

\clearpage
 
\begin{deluxetable}{crrrrrrrrrrr}
\footnotesize
\tablecaption{Summary of the persistent emission characteristics} \label{tbl2}
\tablewidth{0pt}
\tablehead{
\colhead{Model} & \colhead{Model parameter}  
                & \colhead{$N_{\rm H}$ ($10^{22}$~cm$^{-2}$)}  
                & \colhead{reduced $\chi^{2}$~\tablenotemark{b}}
} 
\startdata
BR\tablenotemark{a} : & {\em k}T = $5.77 \pm 0.63$ keV   &  $0.148 \pm 0.453$  &  $0.94$  \\
PL\tablenotemark{a} : & $\Gamma =  2.67 \pm 0.13$  &  $2.27 \pm 0.63$  &  $1.02$  \\
\enddata

 
\tablenotetext{a}{BR = bremsstrahlung,  PL = power law} 
\tablenotetext{b}{144 d.o.f.}

\end{deluxetable}



%
%

\clearpage

%


\begin{thebibliography}{}
\bibitem[Chen, Shrader, \& Livio 1997]{Chen97}
         Chen, W., Shrader, C.R., \& Livio, M. 1997, \apj, 491, 312.
\bibitem[Cocchi et al. 1997a]{Cocc97a} 
         Cocchi, M., et al. 1997a, 
         in {\em "The Active X-ray Sky: results from BeppoSAX and RossiXTE"},
         ed. L. Scarsi, H. Bradt, P. Giommi, \& F. Fiore, Nucl. Phys. B Proc. Suppl., in press
\bibitem[Cocchi et al. 1997b]{Cocc97b} 
         Cocchi, M., et al. 1997b,
         in {\em "The Active X-ray Sky: results from BeppoSAX and RossiXTE"},
         ed. L. Scarsi, H. Bradt, P. Giommi, \& F. Fiore, Nucl. Phys. B Proc. Suppl., in press
\bibitem[Dickey \& Lockman 1990]{Dick90}
         Dickey, J.M., \& Lockman, F.J. 1990, \aapr, 28, 215.
\bibitem[Giles et al. 1996]{Gile96}
         Giles, A.B., et al. 1996, \apj, 469, L25.
\bibitem[Hjellming \& Rupen 1997]{Hjel97} 
         Hjellming, R.M., \& Rupen, M.P. 1997, IAUC 6547.
\bibitem[Hoffman, Marshall, \& Lewin 1978]{Hoff78}
         Hoffman, J.A., Marshall, H.L., \& Lewin, W.H.G. 1978, Nature, 271, 630.
\bibitem[In 't Zand et al. 1998]{Zand98} 
         In 't Zand, J.J.M., et al. 1998, \aap, 331, L25.
\bibitem[Jager et al. 1997]{Jage97} 
         Jager, R., et al. 1997, \aap, 125, 557.
\bibitem[Joss 1978]{Joss78}
         Joss, P.C. 1978, \apj, 225, L123.
\bibitem[Kouveliotou et al. 1996]{Kouv96}
         Kouveliotou, C., et al. 1996, Nature, 379, 799.
\bibitem[Lewin, van Paradijs, \& Taam 1993]{Lewi93} 
         Lewin, W.H.G., van Paradijs, J., \& Taam, R.E. 1993, Space Sci. Rev., 62, 223.
\bibitem[Lewin et al. 1976]{Lewi76}
         Lewin, W.H.G., et al. 1976, \apj, 207, L95.
\bibitem[Lewin, van Paradijs, \& Taam 1995]{Lewi95} 
         Lewin, W.H.G., van Paradijs, J., \& Taam, R.E. 1995, in {\em "X-ray Binaries"},
         ed. W. Lewin, J. van Paradijs, \& E. van den Heuvel,
         Cambridge University Press, Cambridge, p. 175
\bibitem[Marshall et al. 1997]{Mars97} 
         Marshall, F.E., Swank, J.H., Thomas, B., Angelini, L., Valinia, A., \&
         Ebisawa, K. 1997, IAUC 6543.
\bibitem[Stella et al. 1988]{Stel88}
         Stella, L., Haberl, F., Lewin, W.H.G., Parmar, A.N., van Paradijs, J., \&
         White, N.E. 1988, \apj, 324, 379.
\bibitem[Swank et al. 1977]{Swan77}
         Swank, J.H., Becker R.H., Boldt, E.A., Holt, S.S., Pravdo, S.H., \&
         Serlemitsos, P.J. 1977, \apj, 212, L73.
\bibitem[Taam \& Picklum 1979]{Taam79}
         Taam, R.E., \& Picklum, R.E. 1979, \apj, 233, 327.
\bibitem[Tanaka \& Shibazaki 1996]{Tana96} 
         Tanaka, Y., \& Shibazaki, N. 1996, {\em "X-ray novae"}, 
         Annu.Rev.Astron.Astrophys., 34, 607.
\bibitem[van Paradijs 1995]{vanP95} 
         van Paradijs, J. 1995, in {\em "X-ray Binaries"}, 
         ed. W. Lewin, J. van Paradijs, \& E. van den Heuvel,
         Cambridge University Press, Cambridge, p. 536
\bibitem[Voges et al. 1996]{Vogs96} 
         Voges, W., et al. 1996, IAUC 6420.
\bibitem[Woosley \& Taam 1976]{Woos76} 
         Woosley, S.E., \& Taam, R.E. 1976, Nature, 263, 101.
\end{thebibliography}
\end{document}